\documentclass[graybox]{svmult}

\usepackage{mathptmx}       % selects Times Roman as basic font
\usepackage{helvet}         % selects Helvetica as sans-serif font
\usepackage{courier}        % selects Courier as typewriter font
\usepackage{type1cm}        % activate if the above 3 fonts are
\usepackage{makeidx}         % allows index generation
\usepackage{graphicx}        % standard LaTeX graphics tool
\usepackage{multicol}        % used for the two-column index
\usepackage[bottom]{footmisc}% places footnotes at page bottom

\makeindex             % used for the subject index
\begin{document}

\title*{Pulsars as gravitational wave detectors}
% Use \titlerunning{Short Title} for an abbreviated version of
% your contribution title if the original one is too long
\author{George Hobbs}
% Use \authorrunning{Short Title} for an abbreviated version of
% your contribution title if the original one is too long
\institute{George Hobbs \at CSIRO Australia Telescope National Facility, PO Box 76
Epping NSW 1710, Australia, \email{george.hobbs@csiro.au}}
%
% Use the package "url.sty" to avoid
% problems with special characters
% used in your e-mail or web address
%
\maketitle

\abstract{Pulsar timing array projects are carrying out high precision observations of millisecond pulsars with the aim of detecting ultra-low frequency ($\sim 10^{-9}$ to $10^{-8}$\,Hz) gravitational waves.  We show how unambiguous detections of such waves can be obtained by identifying a signal that is correlated between the timing of different pulsars.  Here we describe the ongoing observing projects,  the expected sources of gravitational waves, the processing of the data and the implications of current results.}

\section{Introduction}
\label{sec:1}

The first strong evidence for the existence of gravitational waves (GWs) was obtained by measuring the orbital period decay in the PSR~B1913$+$16 binary pulsar system \cite{ht75a,sb76,tw82}.  Various ground- and space-based detectors have been developed with the aim of making a direct detection of GW signals.  Unfortunately, to date, no such detection has been made. 

Millisecond pulsars are amazingly stable rotators. In many cases a simple model for the pulsar spin-down can be used to predict pulse times-of-arrival (TOAs) with an accuracy and precision of $< 1 \mu$s over many years of observation.  The pulsar timing technique relies on this stability to obtain precise measurements of a pulsar's spin, astrometric and orbital parameters. However, in some cases the pulse TOAs are not exactly as predicted, suggesting that the model and analysis procedure do not parameterise all of the physical effects that are affecting the pulse emission, propagation and detection.  As the effects of GWs are not included in the analysis, the existence of any such waves will induce deviations between the actual and predicted TOAs (commonly known as the  `timing residuals').  Here we describe how these signals may be unambiguously confirmed as being caused by GWs and provide upper bounds on their amplitude.

In \S2 we provide a basic explanation of how pulsar data sets may be used in the search for GWs, \S3 describes current data sets and highlights some aspects of the data that need careful consideration when the data are processed, \S4 lists potential sources of GWs, \S5 describes how data sets may be accessed and \S6 highlights future possibilities.

\section{Using pulsars to search for GWs}

Pulsar observations lead to measurements of TOAs at an observatory.  The \textsc{tempo2} software package \cite{ehm06,hem06} can be used to convert these TOAs to a time of emission by 1) transforming the TOAs to the Solar System barycentre, 2) determining excess propagation delays caused by the interstellar medium and, for binary systems, 3) transforming to the pulsar frame.  The derived time of emission can then be compared with a pulsar model to form the timing residuals.  

The induced pulsar timing residuals caused by a GW signal were first calculated at the end of the 1970s \cite{det79,saz78}.  This early work showed that a GW signal causes a fluctuation in the observed pulse frequency, $\delta \nu/\nu$, which induces pulsar timing residuals at time $t$ from the initial observation as
\begin{equation}
R(t) = - \int_0^t \frac{\delta \nu(t)}{\nu} dt.
\end{equation}
The Dopper shift can be shown to have the form (e.g. \cite{hd83})
\begin{equation}
\frac{\delta \nu}{\nu} = H^{ij}(h_{ij}^e-h_{ij}^p)
\end{equation}
where $h_{ij}^e$ is the GW strain at the Earth at the time of observation, $h_{ij}^p$ the strain at the pulsar when the electromagnetic pulse was emitted and $H^{ij}$ is a geometrical term that depends upon the angle between the Earth, pulsar and GW source.  The GW strains evaluated at the positions of multiple pulsars will be uncorrelated, whereas the component at the Earth will lead to a correlated signal in the timing residuals of all pulsars.  

\begin{figure}[p]
\sidecaption
% Use the relevant command for your figure-insertion program
% to insert the figure file.
% For example, with the graphicx style use
\begin{tabular}{cc}
\includegraphics[scale=.25,angle=-90]{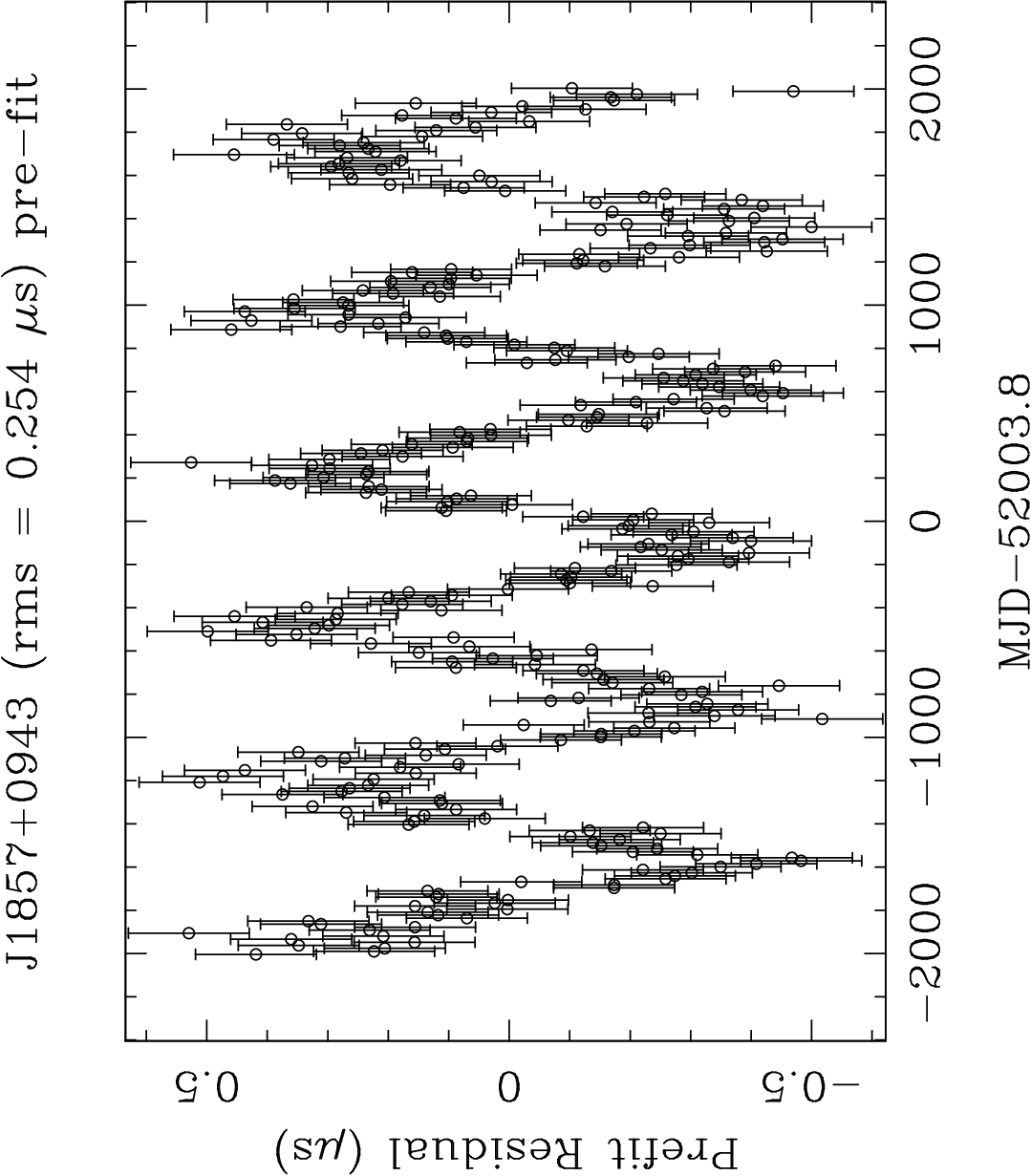} &
\includegraphics[scale=.25,angle=-90]{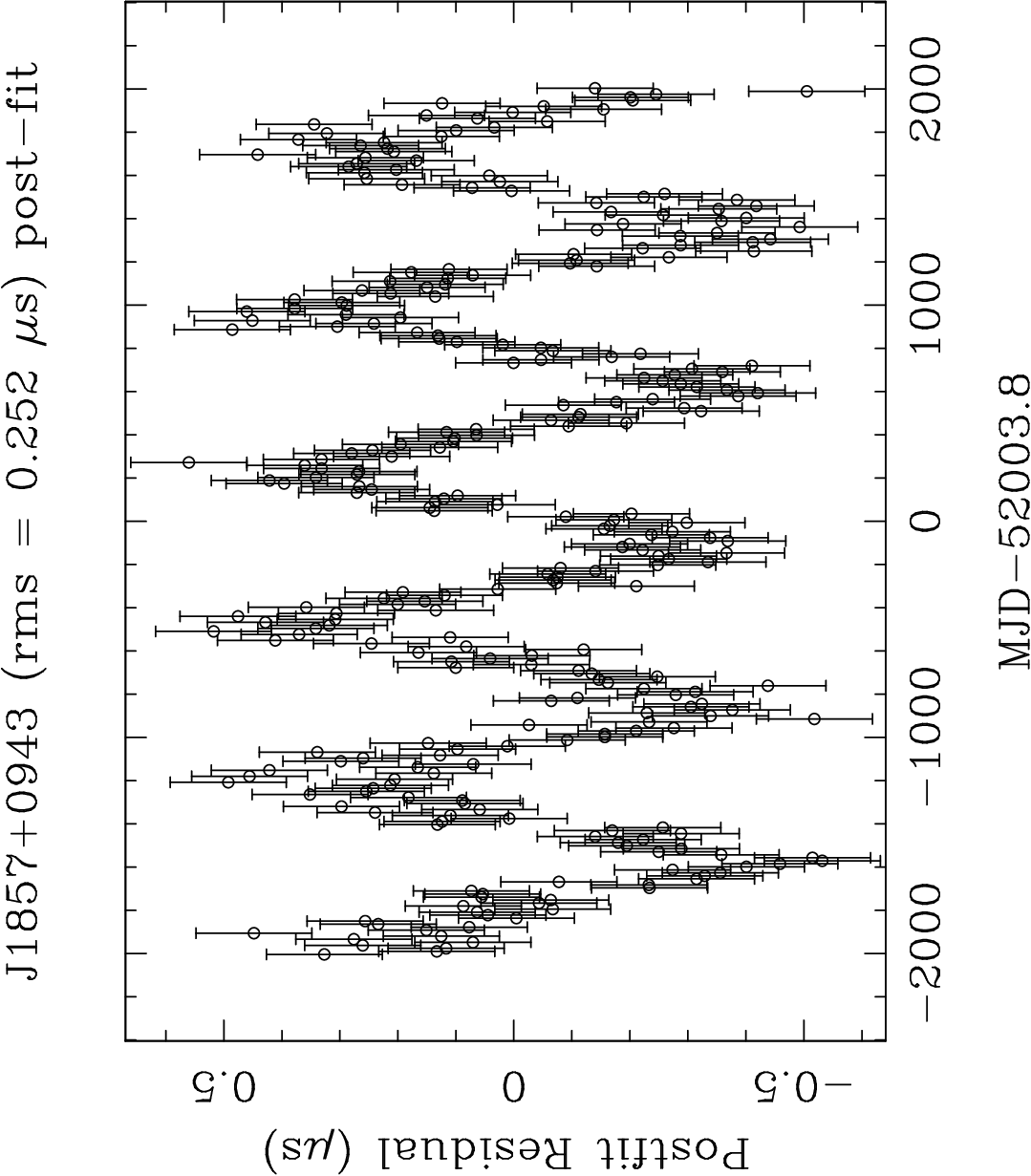} \\
\includegraphics[scale=.25,angle=-90]{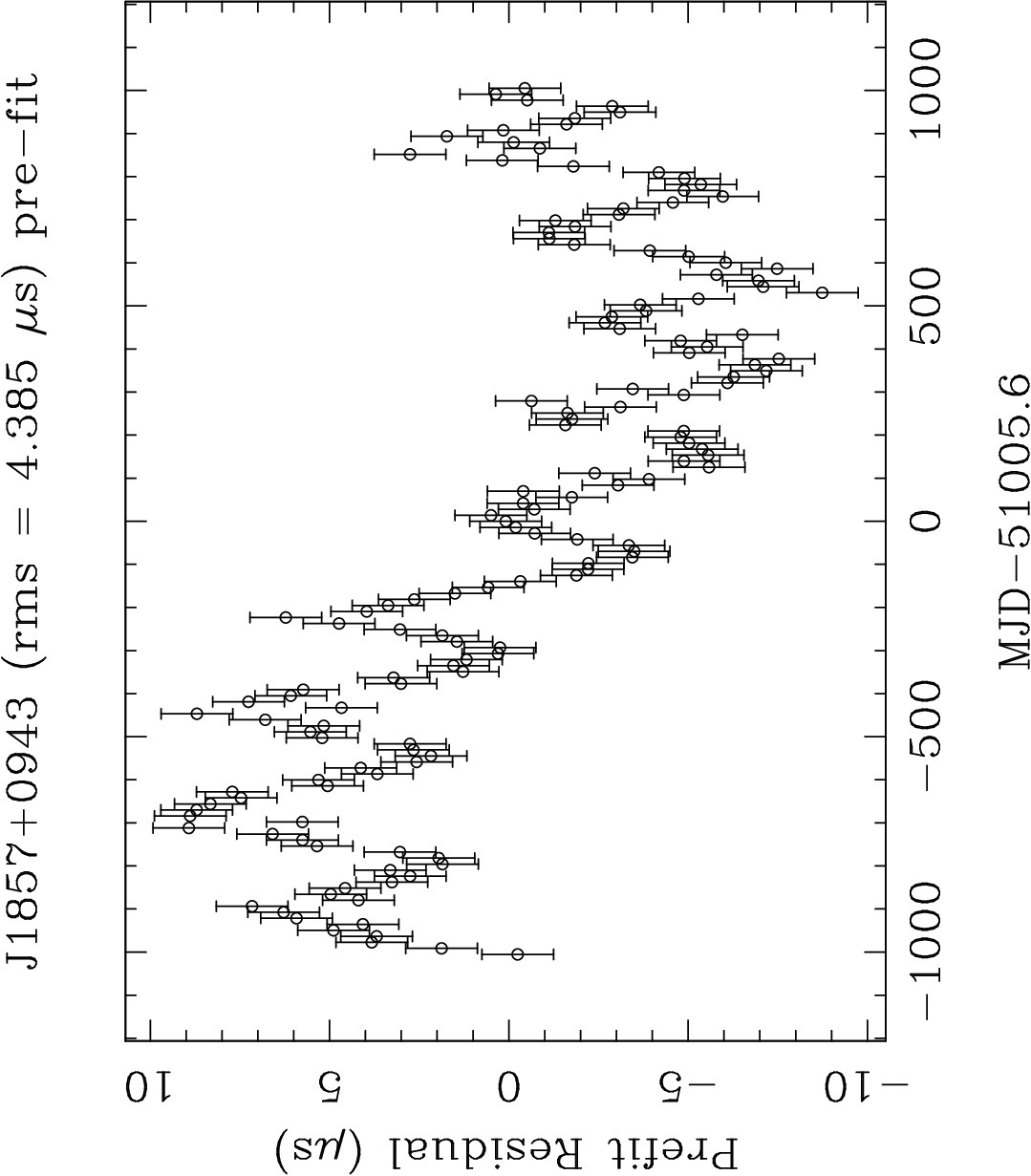} & 
\includegraphics[scale=.25,angle=-90]{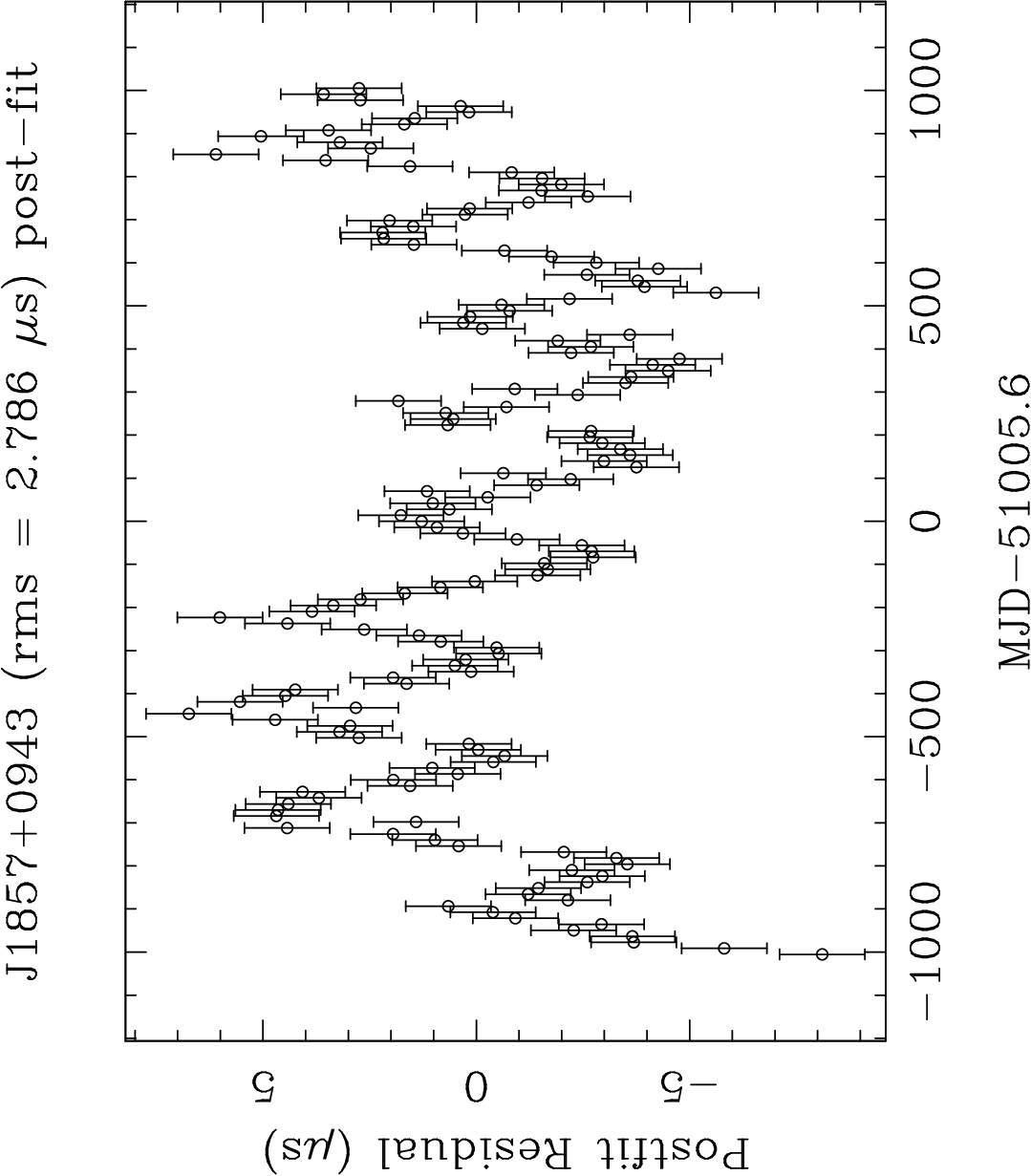} \\
\includegraphics[scale=.25,angle=-90]{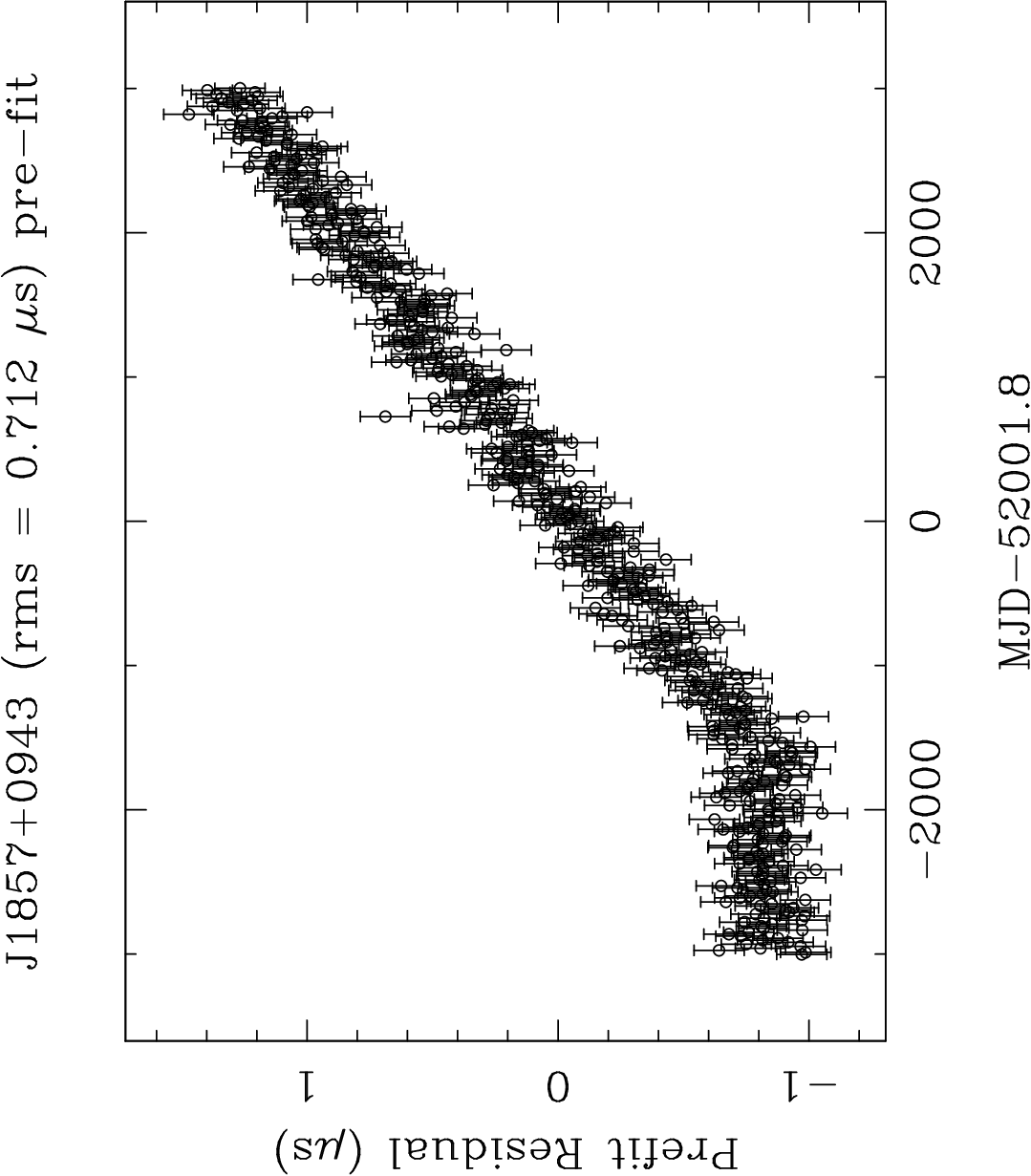} &
\includegraphics[scale=.25,angle=-90]{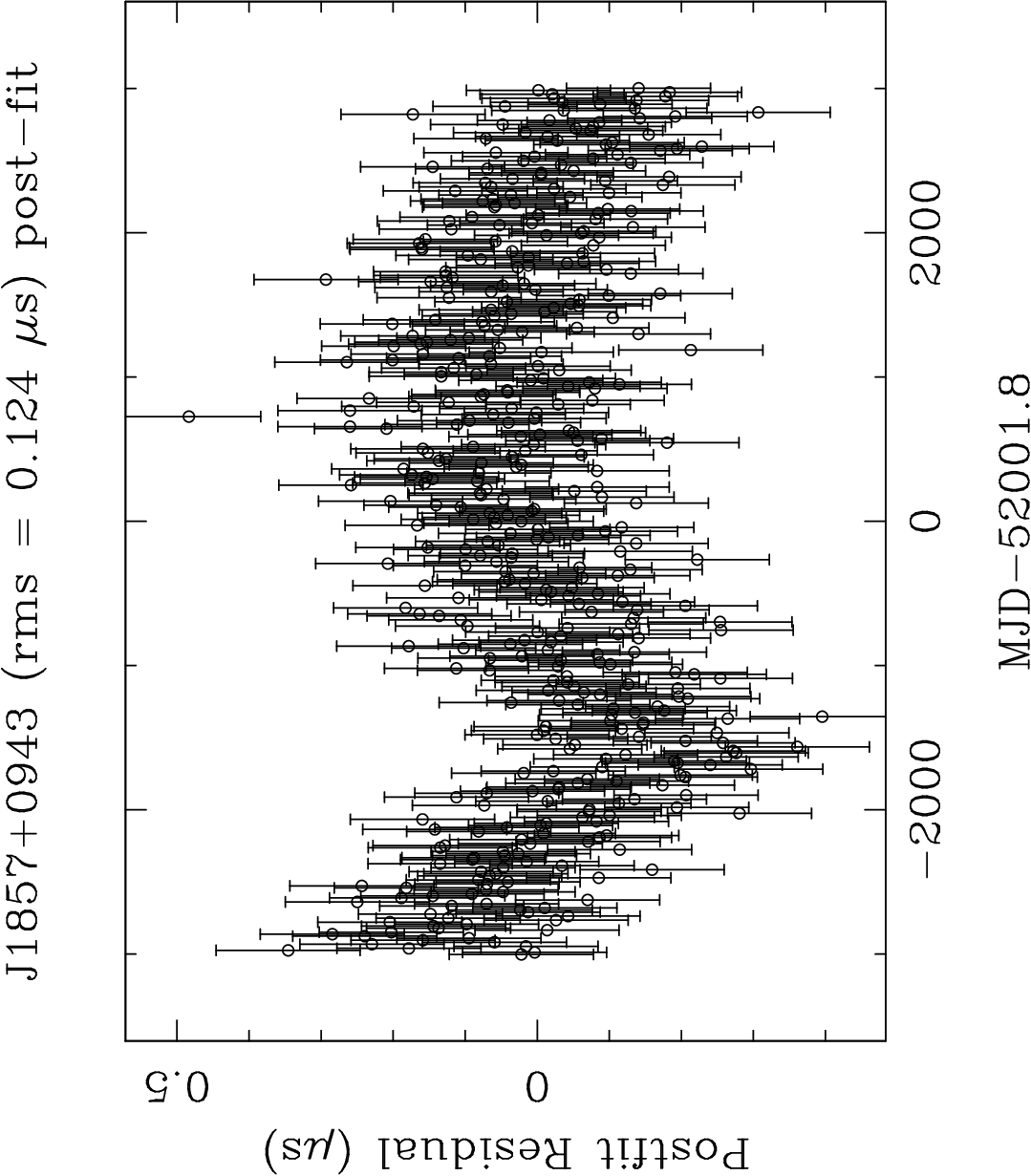} \\
\includegraphics[scale=.25,angle=-90]{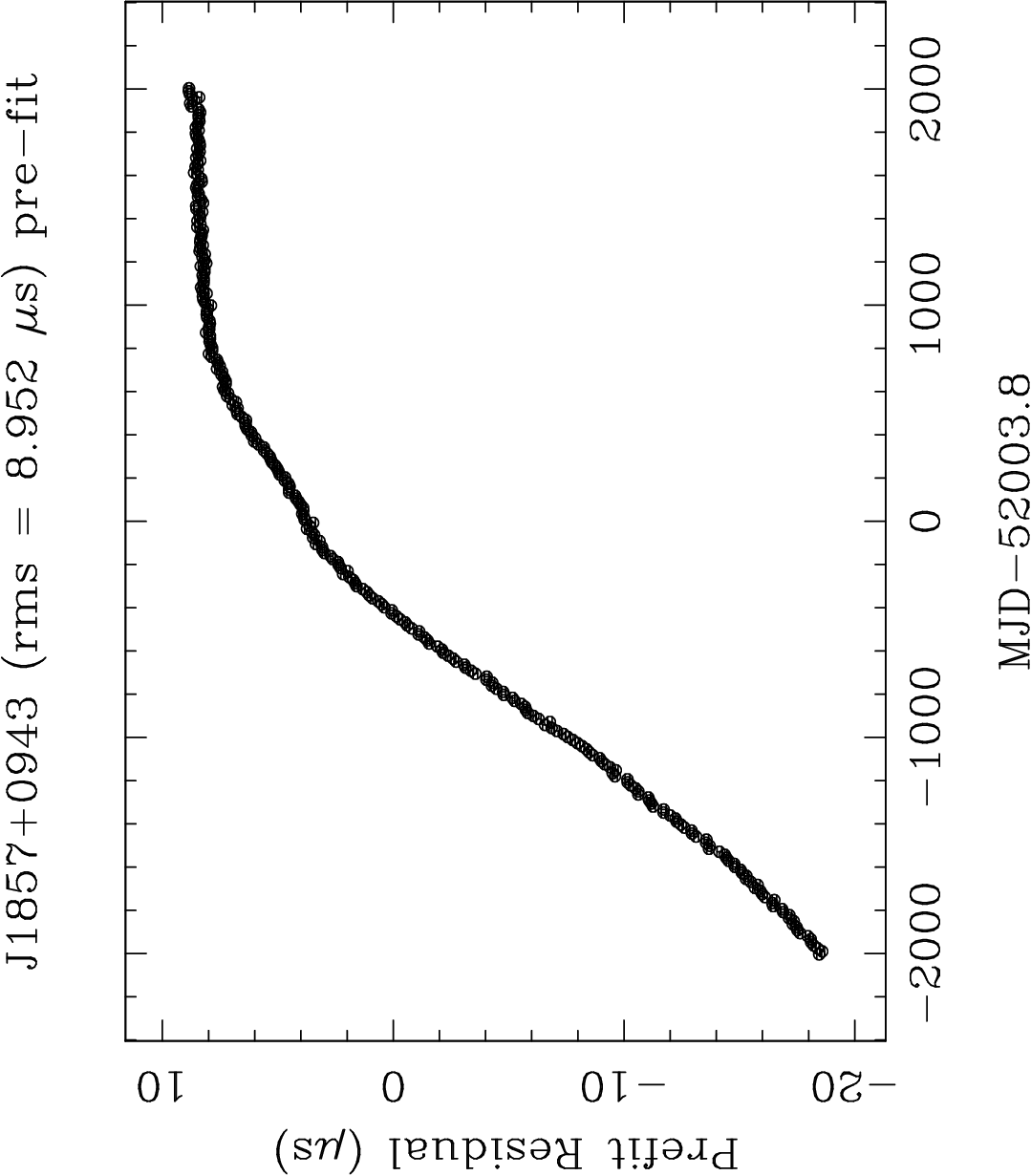} &
\includegraphics[scale=.25,angle=-90]{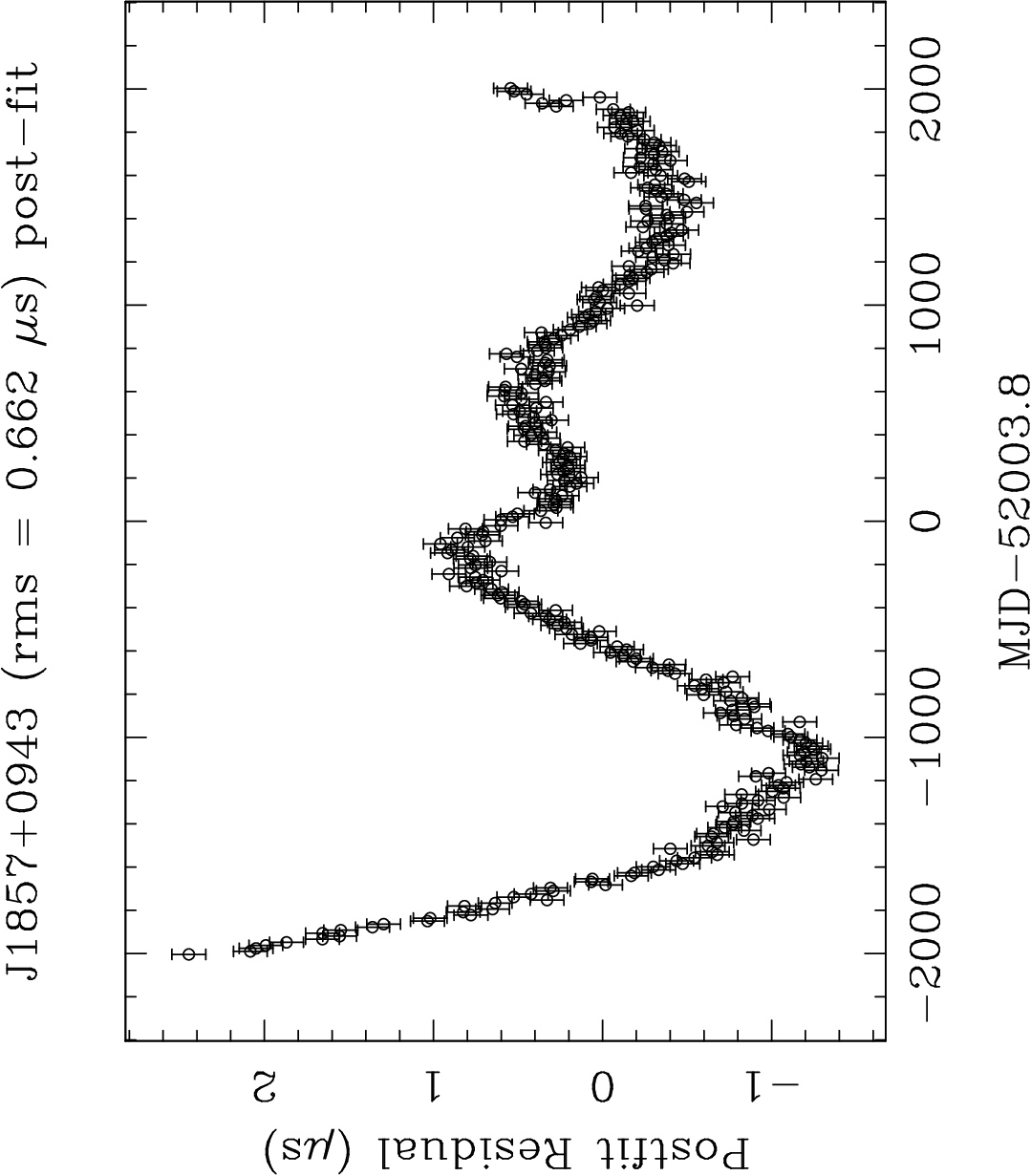} \\
\end{tabular}
%
% If no graphics program available, insert a blank space i.e. use
%\picplace{5cm}{2cm} % Give the correct figure height and width in cm
%
\caption{Simulations of pre-fit (left column) and post-fit (right column) timing residuals. The first row simulates a non-evolving, black-hole binary system defined by $A_+ = A_{\rm X} = 5\times 10^{-14}$ and $\omega = 1\times10^{-7}$.  The second row contains a simulation of the predicted timing residuals from the evolving source 3C66B (see Section~\ref{sec:single}).  The third row contains a GW burst with memory source leading to a change in rotational frequency $\Delta \nu = 1\times 10^{-12}$\,Hz and the bottom row is a realisation of a GW background defined by $\alpha = -2/3$ and $A = 10^{-14}$.}
\label{fg:examples}       % Give a unique label
\end{figure}

% single, non evolving binary source - A+=Ax = 5e-14, omega = 10e-8, gra 0, -gdec - 30, 1855, 14d, 50000, 54000, rms = 100ns - plot1.ps, plot2.ps (fit f0,f1)

% evolve source - as 3C66B .. restricted x-range and rms of 1us

% burst - same as glitch with delta nu = 1e-12

% bkgrd = alpha=-2/3, ngw = 1000, dist=1,gwamp=1e-14

Hobbs et al. (2009 \cite{hjl+09}) demonstrated how pulsar TOAs affected by GWs can be simulated using the \textsc{tempo2} software package. Since the intrinsic pulsar pulse period, spin-down, orbital motion and various astrometric parameters are \emph{a priori} unknown, they need to be determined from the pulsar timing data.  Initial estimates of the pulsar parameters are used to form pre-fit timing residuals.  A least-squares fitting procedure is subsequently used to fit an analytical model to obtain improved pulsar parameters and ``post-fit'' timing residuals (see \cite{hem06}).  In  Figure~\ref{fg:examples} we plot pre- and post-fit timing residuals for a simulated pulsar in the presence of GWs from 1) a single, non evolving binary black-hole system, 2) an evolving binary black-hole system, 3) a burst GW source and 4)  an isotropic stochastic background of GWs.

% For figures use
%
\begin{figure}[t]
\sidecaption
% Use the relevant command for your figure-insertion program
% to insert the figure file.
% For example, with the graphicx style use
\includegraphics[scale=.45,angle=-90]{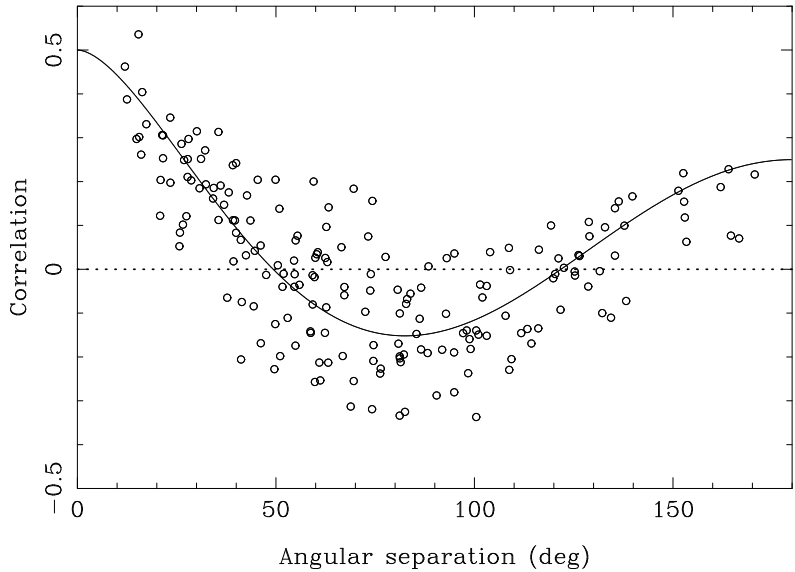}
%
% If no graphics program available, insert a blank space i.e. use
%\picplace{5cm}{2cm} % Give the correct figure height and width in cm
%
\caption{The expected correlation in the timing residuals of pairs of pulsars as a function of angular separation for an isotropic GW background (solid line).  The points correspond to simulated pulsar data sets in the presence of a GW background with power-law index $\alpha = +3/2$ and amplitude $A = 0.01$; Figure from Hobbs et al. (2009, \cite{hjl+09}).}
\label{fg:hdcorr}       % Give a unique label
\end{figure}

It is not possible to determine the exact origin of the timing residuals for a single pulsar data set.  Any observed residuals may have been caused by, for example, irregularities in terrestrial time standards \cite{rod08}, errors in the planetary ephemeris, irregular spin-down of the pulsar \cite{hlk10}, calibration effects (e.g. \cite{van06}) or GWs.  These effects can only be distinguished by searching for correlations in the timing residuals of multiple pulsars.  For instance, residuals caused by the irregular spin-down of one pulsar will be uncorrelated with the residuals observed for a different pulsar.  Irregularities in terrestrial time standards will lead to correlated residuals for all pulsars (this is only true for data sets which have the same data span; see Section~\ref{sec:current}).  For an isotropic, stochastic GW background, the GW strain at each pulsar will be uncorrelated, but the GW strain at the Earth provides a common signal. The common signal for such a background was determined by Hellings \& Downs (1983)\nocite{hd83} and is reproduced in Figure~\ref{fg:hdcorr}.  More recently it has been shown that approximately 20 pulsars are necessary, timed with an rms timing residual of $\sim$100\,ns or better over five years with observations every week, in order to make a significant detection of a possible GW background signal \cite{jhlm05}.

\begin{figure}[t]
\sidecaption
% Use the relevant command for your figure-insertion program
% to insert the figure file.
% For example, with the graphicx style use
\includegraphics[scale=.45,angle=-90]{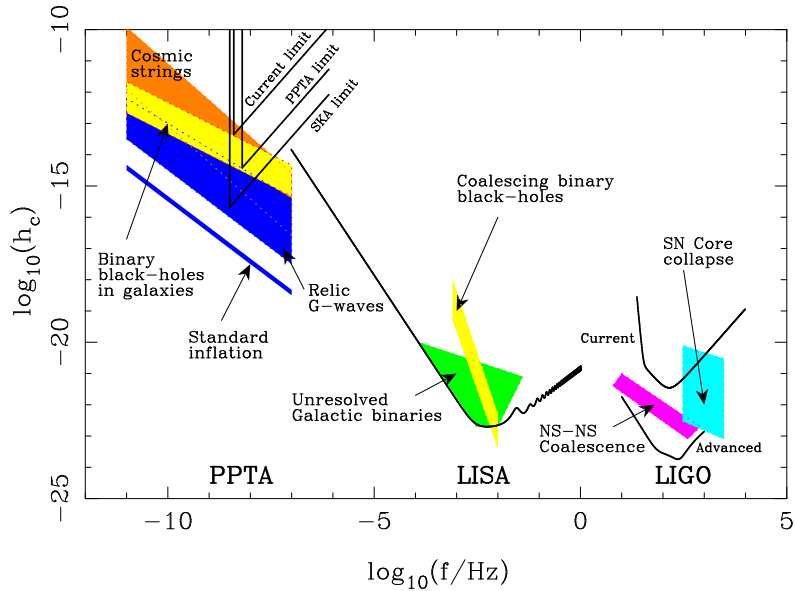}
%
% If no graphics program available, insert a blank space i.e. use
%\picplace{5cm}{2cm} % Give the correct figure height and width in cm
%
\caption{Characteristic strain sensitivity for existing and proposed GW detectors as a function of GW frequency. Predicted signal levels from various astrophysical GW backgrounds are shown. The current limit is that published by Jenet et al. (2006).  The `PPTA limit' is a predicted limit if the PPTA project achieves its design sensitivity.  The SKA limit is a prediction of the sensitivity of the planned Square Kilometre Array telescope.}
\label{fg:sens}       % Give a unique label
\end{figure}

Pulsar timing experiments are sensitive to GW signals in the ultra-low frequency band (f $\sim 10^{-9}$ to $10^{-8}$\,Hz) as individual pulsars are typically only observed once every few weeks, have maximum data spans of years to decades and require the fitting of a pulsar timing model.  The pulsar timing method is therefore complementary to other GW detection methods such as the Laser Interferometer Space Antenna (LISA)\footnote{http://lisa.nasa.gov} and ground based interferometer systems (such as the Laser Interferometer Gravitational Wave Observatory, LIGO\footnote{http://www.ligo.caltech.edu/advLIGO/}), which are sensitive to high frequency GWs. In Figure~\ref{fg:sens} we plot the sensitivity of the pulsar timing experiments, LISA and LIGO as well as some expected GW sources.\footnote{Various authors e.g. \cite{kop97,psh09} have indicated that it may be possible to limit the existence of GWs with even lower frequencies ($10^{-12}$ to $10^{-8}$\,Hz) where the minimum sensitive GW frequency is determined by the distance to the pulsar.  However, the methods required to achieve this sensitivity are different to those described in this paper.  We also note that a few other methods exist that can limit the ultra-low frequency GW background.  One example is reported by Gwinn et al. (1997)\nocite{gep+97} who obtained an upper bound on the background amplitude by analysing limits on proper motions of quasars.}  Pulsar timing projects that aim to obtain data sets on the most stable millisecond pulsars with the aim of searching for correlated timing residuals are known as pulsar `timing arrays'.  The first attempts to undertake such projects are described by Romani (1989)\nocite{rom89} and Foster \& Backer (1990)\nocite{fb90}.

\section{Current data sets}\label{sec:current}

\begin{figure}[t]
\sidecaption
% Use the relevant command for your figure-insertion program
% to insert the figure file.
% For example, with the graphicx style use
\includegraphics[scale=.45,angle=-90]{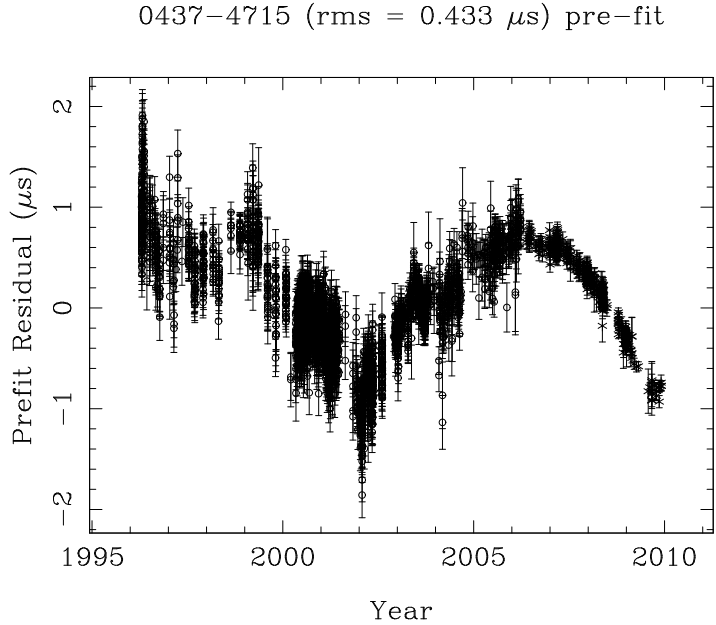}
%
% If no graphics program available, insert a blank space i.e. use
%\picplace{5cm}{2cm} % Give the correct figure height and width in cm
%
\caption{The timing residuals for PSR~J0437$-$4715 over 14 years.}
\label{fg:0437}       % Give a unique label
\end{figure}

The International Pulsar Timing Array (IPTA) project has the main aim of detecting GW signals using pulsar observations and is a collaboration between three separate projects. The European project (EPTA \cite{fvb+10}) currently obtains data using the Effelsberg, Jodrell Bank, Nan\c cay and Westerbork telescopes (a new telescope in Sardinia is currently under construction and, when commissioned, will be used as part of the EPTA).  The North American project (NANOGrav \cite{jfl+09}) uses the Arecibo and Green Bank telescopes and the Parkes project (PPTA; e.g. \cite{hbb+09,man10,vbb+10}) uses the Parkes radio telescope in Australia.  In total approximately 37 different pulsars are being observed by these projects\footnote{The exact number of pulsars varies as new pulsars are discovered and added to the IPTA sample. Radio pulsar surveys are being carried out (Keith et al., submitted to MNRAS,\cite{cfl+06}) and the Fermi gamma-ray spacecraft, has recently discovered numerous previously unidentified gamma-ray sources.  Many of these have now been shown to be millisecond pulsars (P. Ray, these proceedings) some of which are likely soon to be included in the IPTA.} (see \cite{haa+10}).   Some of the most precise timing residuals have been obtained for the brightest millisecond pulsar, PSR J0437$-$4715.  The rms timing residuals over $\sim 1$\,yr obtained with the Parkes telescope at an observing frequency close to $3$\,GHz is $\sim$60\,ns and some individual TOA uncertainties are $\sim$30\,ns.  Achieving such low rms timing residuals relies on removing the effects of dispersion measure variations \cite{yhc+07}, precisely calibrating the data \cite{vmjr09} and determining many post-Keplerian binary parameters \cite{vbv+08}. However, over longer data spans clear irregularities are observed in the timing residuals (see Figure~\ref{fg:0437}; R. Manchester, private communication).  These residuals differ from those published by Verbiest et al. (2008)\nocite{vbv+08} as 1) new data have been added between the years 2006 and 2010 and 2) the time offsets between different backend systems at the observatory have been measured and subsequently not included in the fitting procedures.  The cause of the observed variations is not currently known. 

Pulsars exhibit two main types of timing irregularity: `glitches' in which the pulsar's rotation rate suddenly increases before undergoing a period of relaxation and `timing noise' which consists of low-frequency features in the timing residuals.   Hobbs et al. (2010)\nocite{hlk10} analysed a sample 366 pulsars and showed that the timing noise, in general, consisted of quasi-periodic oscillations with periodicities of years to decades.  However, the ``amount'' of timing noise decreases with characteristic age and millisecond pulsars, which are used in timing array experiments, are typically extremely stable.  In a few cases, timing noise and glitches are clearly seen in the residuals for millisecond pulsars.  For instance, the timing residuals for PSR~J1939$+$2134 have long been known (e.g. \cite{ktr94}) to be dominated by timing noise.    The first glitch in a millisecond pulsar was found in PSR~B1821$-$24 which is located in the globular cluster M28 \cite{cb04}.  The glitch was shown to follow the main characteristics of glitches seen in the slower pulsars.  Verbiest et al. (2009)\nocite{vbc+09} studied the timing stability for 20 millisecond pulsars included in the PPTA project, over time scales up to 10\,yr. Apart from PSR~J1939$+$2134 it was concluded that the timing of most of the other pulsars is stable enough for GW detection on decadal time scales.

\begin{figure}[t]
\sidecaption
% Use the relevant command for your figure-insertion program
% to insert the figure file.
% For example, with the graphicx style use
\includegraphics[scale=.25,angle=-90]{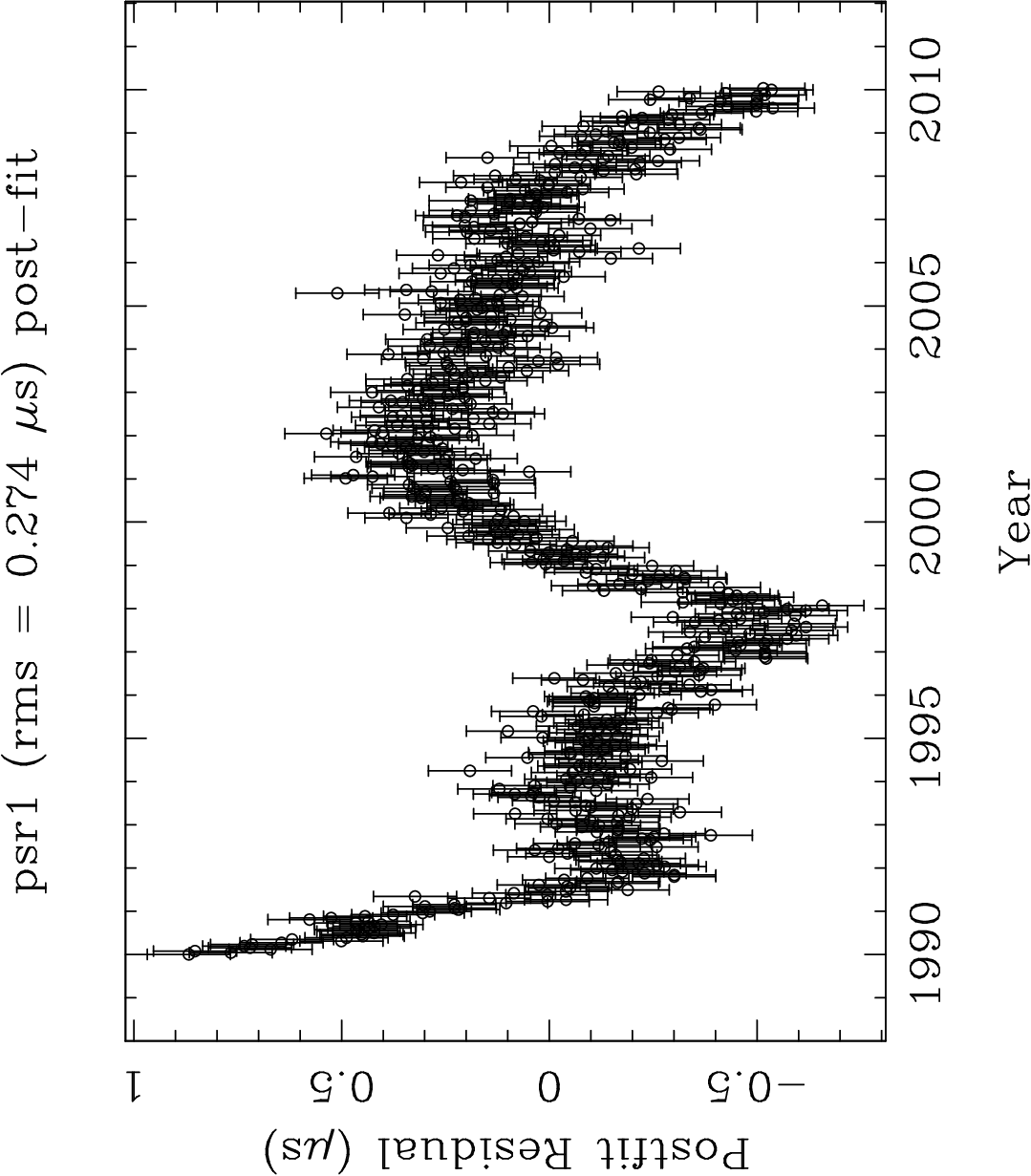}
\includegraphics[scale=.25,angle=-90]{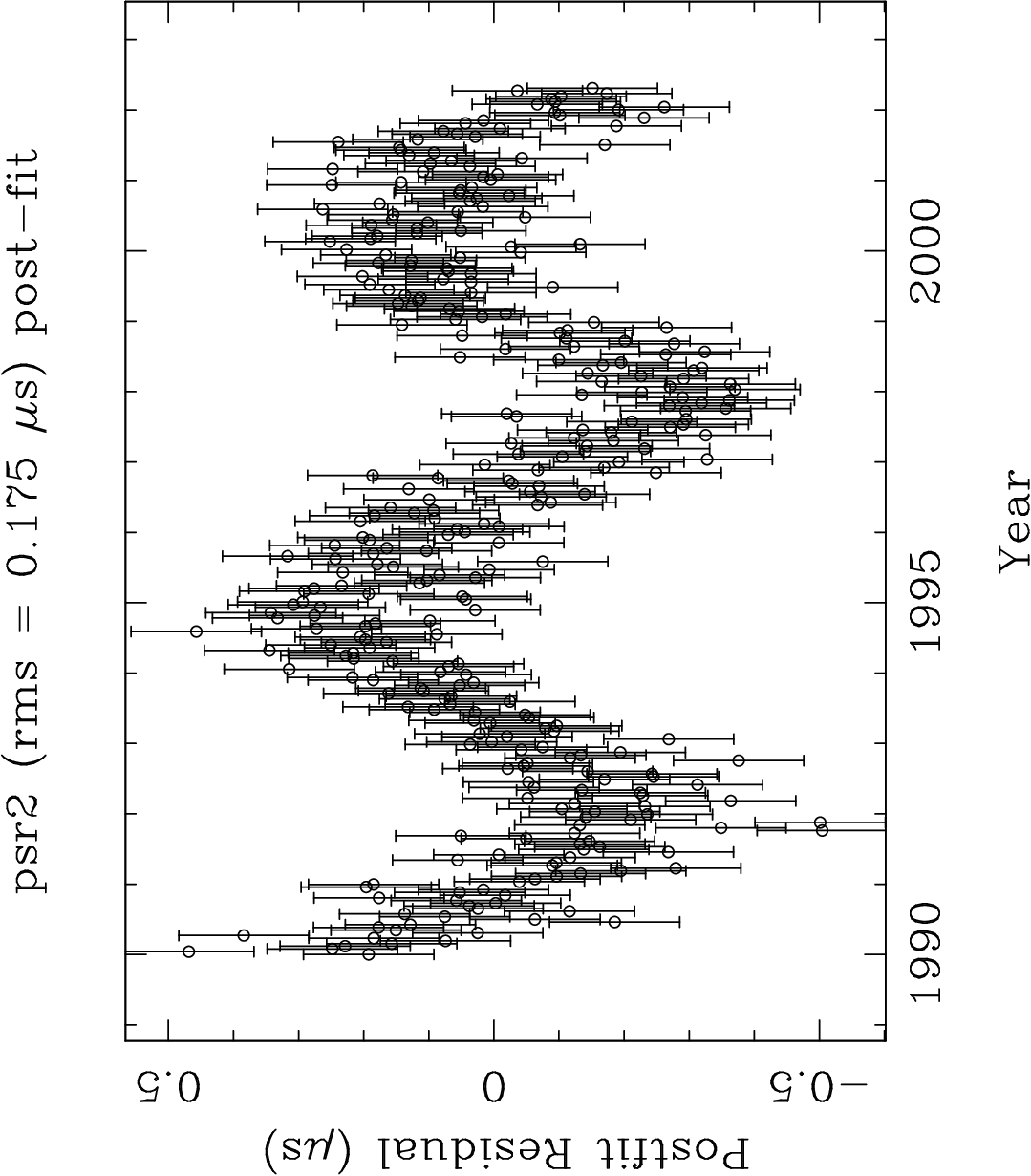}
\includegraphics[scale=.25,angle=-90]{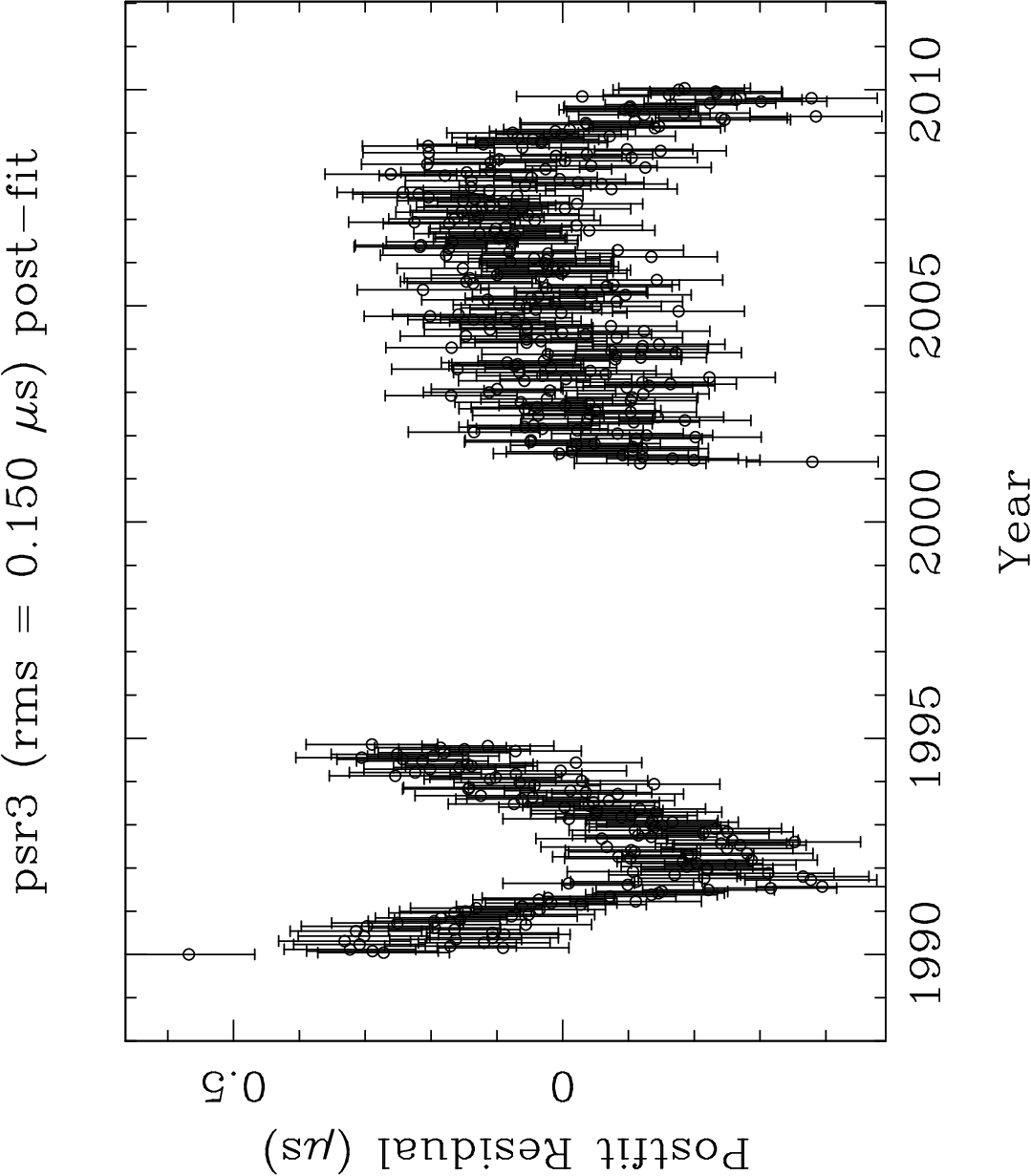}
%
% If no graphics program available, insert a blank space i.e. use
%\picplace{5cm}{2cm} % Give the correct figure height and width in cm
%
\caption{Post-fit timing residuals for three simulated pulsars.  Each pulsar data set consists of white noise plus a clock irregularity that is the same for each pulsar.  The post-fit residuals are not correlated because of model fitting and because each data set has a different data span.}
\label{fg:clock}       % Give a unique label
\end{figure}

The basic method to detect a GW background signal in pulsar data is to search for correlated timing residuals between different pairs of pulsars. However, we note that these correlations only hold in the  case where all pulsar data sets are processed using the same fitting procedure and have the same data span. In Figure~\ref{fg:clock} post-fit timing residuals are shown that include simulated clock irregularities. Even though the same signal has been added to each data set, the residuals are not highly correlated because of the differing data spans and the model fits that have been carried out.  The following need to be taken into account when attempting to detect or limit GW signals in pulsar timing data:

\begin{itemize}
\item Pulsar timing residuals are irregularly sampled, have large gaps and different pulsars have different sampling and data spans.

\item The uncertainty on each TOA for a given pulsar can vary by an order-of-magnitude because of interstellar scintillation, different observing durations for different observations and when combining data sets obtained from different observatories.  For an unknown reason it is often found that the TOA uncertainties are under- or over-estimated (by factors of $\sim$2).

\item Timing residuals are always related to a pulsar timing model for which various fits have been carried out to determine the pulsar's spin, astrometric and orbital parameters.

\item Arbitrary offsets can exist between TOAs measured with different instruments or at different observatories.

\item The timing residuals may be affected by timing noise or glitch events.

\end{itemize}

\section{Potential sources of gravitational waves}\label{sec:single}

Theoretical models predict that pulsar timing experiments are sensitive to individual, periodic sources of GWs, burst sources and a background of GWs. This section describes the various possible sources that have been discussed in the literature.

\subsection{Single sources}

GWs are generated by the acceleration of massive objects.  For a two-body orbital system the GW frequency is twice the orbital frequency implying that supermassive binary black-hole systems can produce GW signals that are strong enough, and are in the correct frequency regime, to be detectable by pulsar timing experiments. An estimate of the timing residual for a massive black-hole binary system can be obtained from \cite{jfl+09}:
\begin{equation}
t \sim 10{\rm ns}\left(\frac{1{\rm Gpc}}{d}\right) \left(\frac{M}{10^9 {\rm M}_\odot}\right)^{5/3} \left(\frac{10^{-7}\rm{Hz}}{f}\right)^{1/3}
\end{equation}
where $d$ is the distance to the black-hole binary which has a total mass of  $M/(1+z)$ and emits GWs  at frequency $f$. 

Sesana, Vecchio \& Volonteri (2009)\nocite{svv09} showed that it is likely that the maximum timing residuals induced by such a system will be in the $\sim 5 - 50$\,ns range. Any such source is likely to be from a massive system (with a chirp mass $M_c > 5 \times 10^8$\,M$_\odot$) and at a redshift between $0.2 < z < 1.5$.  More recently Sesana \& Vecchio (2010)\nocite{sv10} showed for a ``fiducial'' timing array project containing 100 pulsars uniformly distributed in the sky that, if a source is detected with a signal to noise ratio of 10, the source position could be determined to within $\sim 40$\,deg$^2$ with a fractional error on the signal amplitude around 30\% and the source inclination and polarisation angles could be recovered to within $\sim 0.3$\,rad.

Unfortunately there has not yet been any clear detection of a supermassive binary black hole system where the black holes are close enough, and massive enough, for the system to be emitting detectable GW emission.  Rodriguez et al. (2006)\nocite{rtz+06} discovered a system in the radio galaxy 0402$+$379, which has a projected separation between the two black holes of just 7.3 pc.  However, this is still too wide to be emitting detectable GWs. Sudou et al. (2003)\nocite{simt03} reported the possible detection of a supermassive black hole binary in the radio galaxy 3C 66B and provided the orbital parameters for the postulated system.  Jenet et al. (2004)\nocite{jllw04} showed that such a system should have produced detectable GW emission and managed to rule out its existence with high confidence.  Lommen \& Backer (2001)\nocite{lb01} unsuccessfully searched for  GW emission from Sagittarius A$^*$ which had also been suggested to be part of a binary system. Sillanpaa et al. (1996)\nocite{stp+96} have identified a candidate supermassive black hole binary system in the blazar OJ287 with member masses of $1.3 \times 10^8$\,M$_\odot$ and $1.8 \times 10^{10}$\,M$_\odot$.  However, Yardley et al. (2010)\nocite{yhj+10} showed that this system will require the sensitivity of future telescopes (such as the Square Kilometre Array, SKA) for detection.  Searches for more supermassive binary black hole candidates are ongoing.  Wen, Liu \& Han (2009)\nocite{wlh09} selected 1209 pairs of galaxies from the Sloan Digital Sky Survey and searched for features indicative of merging events.   More recently Burke-Spolaor (submitted to MNRAS) used archival VLBI data to search 3114 radio-luminous active galactic nuclear for binary supermassive black holes.  Only one such source (the same as that previously discovered by Rodriguez et al. 2006) was detected.

\begin{figure}[t]
\sidecaption
% Use the relevant command for your figure-insertion program
% to insert the figure file.
% For example, with the graphicx style use
\includegraphics[scale=.45,angle=-90]{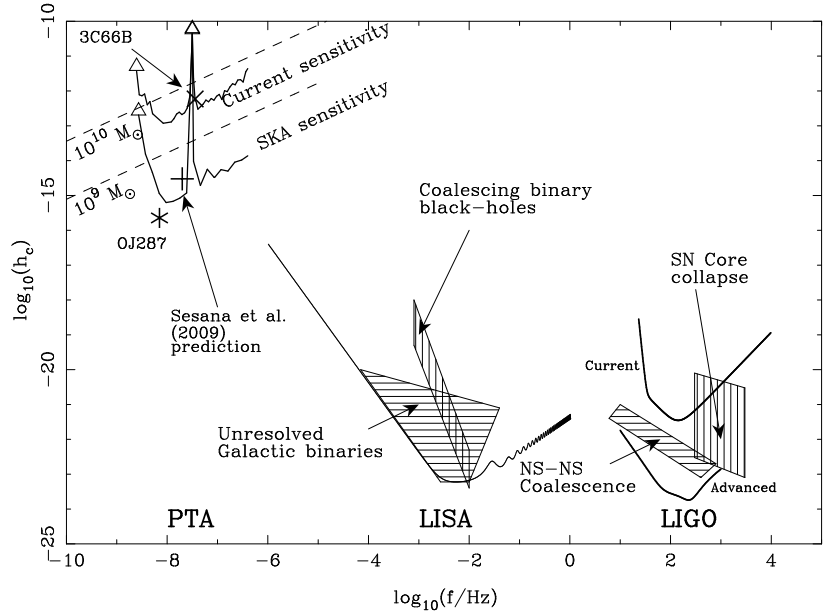}
%
% If no graphics program available, insert a blank space i.e. use
%\picplace{5cm}{2cm} % Give the correct figure height and width in cm
%
\caption{Figure from Yardley et al. (2010) showing the sensitivity of current and future GW observatories to individual, periodic GW sources as a function of the GW frequency.}
\label{fg:gwsingle}       % Give a unique label
\end{figure}

Wen et al. (submitted to ApJ) described the basic methods in which existing pulsar timing observations can be used to place constraints on the coalescence rate of binary supermassive black holes in the Universe. Yardley et al. (2010)\nocite{yhj+10} determined the sensitivity of the Verbiest et al. (2008, 2009) data sets to individual GW sources and obtained a sky-averaged constraint on the merger rate of nearby ($z < 0.6$) black-hole binaries in the early phases of coalescence with a chirp mass of $10^{10}$\,M$_\odot$ of less than one merger every seven years.  This work is summarised in Figure~\ref{fg:gwsingle} which plots the current sensitivity to single sources and predicted future sensitivity (with the Square Kilometre Array, SKA, telescope).

\subsection{Burst sources}

Sources of detectable burst GW emission include 1) the formation of supermassive black holes  which leads to a day-long burst of radiation \cite{tb76}, 2) highly eccentric supermassive black hole binaries \cite{en07}, 3) close encounters of massive objects \cite{kgm06} and 4) cosmic string cusps \cite{dv01}.   Pshirkov \& Tuntsov (2010)\nocite{pt10} constrained the cosmological density of cosmic string loops using photometry and from pulsar timing showing that the pulsar timing data provides the most stringent constraints on the abundance of light strings. 

Seto (2009)\nocite{set09}, Pollney \& Reisswig (2010)\nocite{pr10}, van Haasteren \& Levin (2010)\nocite{vl10} and Pshirkov, Baskaran \& Postnov (2010)\nocite{pbp10} all considered GW bursts ``with memory''.  In general such events consist of a rise in the GW field, followed by an oscillatory behaviour for a few cycles and finally convergence to a non-zero value.  The permanent change is known as the burst's ``memory''.  Such events can occur during close encounters of massive bodies on hyperbolic trajectories or in an asymmetric supernova. Pshirkov, Baskaran \& Postnov (2010) showed that one such event could be detected from distances of up to 1 Gpc (for the case of equal mass supermassive binary black hole systems of $M = 10^8$\,M$_\odot$).  A similar event can occur if the line of sight to a pulsar passes a cosmic string (Pshirkov \& Tuntsov 2010). Such an event would instantaneously change the apparent frequency, $\nu$, of the pulsar by a small amount $\Delta \nu$ in the same way as that of a GW burst with memory although the memory phenomenon would apply to all observed pulsars whereas this line-of-sight effect is specific to an individual pulsar.   

Amaro-Seoane et al. (2010)\nocite{ash+10} discussed the possibility that supermassive black holes do not coalesce before the merger with a third galaxy.  This leads to orbits with high eccentricity leading to an intense burst of GWs.  They showed that, for reasonable supermassive black-hole evolution models, a few bursts will induce timing residuals $> 1$\,ns, however, if pulsar timing were sensitive at the nanosecond level, most of these bursts would be undetectable because of confusion with the GW background.

A detection of a burst source using pulsar timing would provide the ability to localise the position of the source with the hope of identifying an electromagnetic counterpart.  The angular resolution of a given timing array experiment depends upon the number of pulsars in the array, the timing precision and the distribution of the pulsars on the sky.  

\subsection{Stochastic background}

\begin{figure}[t]
\sidecaption
% Use the relevant command for your figure-insertion program
% to insert the figure file.
% For example, with the graphicx style use
\includegraphics[scale=.45,angle=-90]{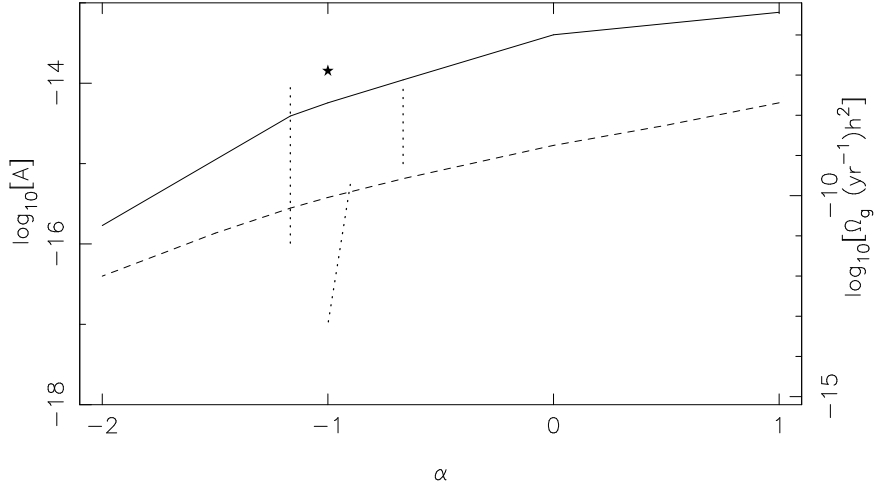}
%
% If no graphics program available, insert a blank space i.e. use
%\picplace{5cm}{2cm} % Give the correct figure height and width in cm
%
\caption{Upper bounds from Jenet et al. (2006) for stochastic backgrounds with various values of the GW spectral exponent $\alpha$ (solid line).  The dotted lines indicate regions predicted by various GW background models. The dashed line indicates potential future limits from the PPTA project.  The star symbol provides the upper bound given by Kaspi, Taylor \& Ryba (1994).  Figure from Jenet et al. (2006).}
\label{fg:aupper}       % Give a unique label
\end{figure}

A stochastic background of GWs is expected from the early phases of coalescence for supermassive black holes \cite{rr95a,jb03,wl03a,eins04,shmv04}, cosmic strings or relic GWs from the big bang \cite{mag00}. In most models, the GW strain spectrum, $h_c(f)$, is represented by a power-law in the GW frequency, $f$,
\begin{equation}
h_c(f) = A\left(\frac{f}{{\rm yr^{-1}}}\right)^\alpha
\end{equation}
where the spectral exponent, $\alpha = -2/3$, $-1$ and $-7/6$ for likely GW backgrounds caused by coalescing black hole binaries, cosmic strings and relic GWs respectively.  The energy density of the background per unit logarithmic frequency interval can be written as
\begin{equation}
\Omega_{\rm GW}(f) = \frac{2}{3}\frac{\pi^2}{H_0^2}f^2h_c(f)^2
\end{equation}
where $H_0$ is the Hubble constant.  Sesana, Vecchio \& Colacino (2008)\nocite{svc08} showed that the expected form of the stochastic background is not simply $h_c \propto f^{-2/3}$, but the frequency dependence becomes steeper above $\sim 10^{-8}$\,Hz.  They showed that the major contributors to the background come from massive $>10 ^{8}$M$_\odot$ binary black hole systems that are relatively nearby ($z < 2$).

In contrast to backgrounds formed from supermassive black-hole binary systems, Saito \& Yokoyama (2009)\nocite{sy09} showed that the formation of intermediate-mass ($\sim$600\,M$_\odot$) black holes at the end of the inflationary era will lead to a potentially detectable background of GWs.  Current pulsar upper bounds already limit the process for the formation of such black holes. 

Both ``kinks'' and ``cusps'' on cosmic string loops can also contribute to a background of GWs.  Olmez, Mandic \& Siemens (2010)\nocite{oms10} showed that both phenomena contribute at the same order to the background.  The most recent constraints, and their implications, on the cosmic string tension using pulsar timing experiments have been provided by Battye \& Moss (2010)\nocite{bm10}.  

A background of GW radiation from the early universe is also expected, if detected, to provide a unique view of the physics of the very early universe (e.g. \cite{gri05}).  Boyle \& Buonanno (2007)\nocite{bb07} showed that combinations of cosmic microwave background experiments and GW detection experiments lead to strong constraints on the existence and properties of various energy components that may have dominated the universe at the end of the inflationary era.

One of the first limits on a GW background was placed by Stinebring et al. (1990)\nocite{srtr90}.  Kaspi, Taylor \& Ryba (1994)\nocite{ktr94} used observations of PSRs~J1857+0943 and J1939+2134 from the Arecibo observatory to provide an upper bound on the GW background. Their data sets were made publically available allowing other algorithms to be applied to their data sets \cite{mzvl96,jhv+07}.   Jenet et al. (2006) developed a method that took into account all the effects of fitting a pulsar timing model to data sets of different data spans and applied the technique to carefully selected PPTA data sets as the technique can only be applied to data sets that are statistically ``white''.  Van Haasteren et al. (2008)\nocite{vlml08} developed a Bayesian algorithm which has been applied to EPTA data sets \cite{fvb+10}.  Anholm et al. (2009)\nocite{abc+09} developed an independent method which currently does not address some issues relating to real pulsar data.  Work is ongoing to develop a method that correctly detects the GW background or limits its amplitude and can easily be applied to the IPTA data sets.

The most stringent limits published to date on the amplitude of a background caused by these various phenomena were obtained using observations from the PPTA project combined with archival data from the Arecibo telescope \cite{jhv+07}.  These upper bounds on $A$ are shown in Figure~\ref{fg:aupper}.  These results were shown to 1) constrain the merger rate of supermassive black hole binary systems at high redshift, 2) rule out some relationships between the black hole mass and the galactic halo mass, 3) constrain the rate of expansion in the inflationary era and 4) provide an upper bound on the dimensionless tension of a cosmic string background.

\section{Accessing pulsar data sets}

The IPTA pulsar data sets contain some of the most precisely measured observations yet made of radio pulsars.  These data sets can be used to search for GW sources and for numerous other purposes. Details of the only currently publically available data sets are provided in Kaspi, Taylor \& Ryba (1994)\nocite{ktr94}.  Hobbs et al. (2009)\nocite{hjl+09} provided a sample of simulated data sets for comparison of GW detection algorithms that are available for download.  In the near future it is expected that all observations from Parkes Observatory (after an embargo period of 18 months) will be available for download providing access to most PPTA data.  A few data sets relating to the PPTA project are also available as part of the PULSE@Parkes outreach project\cite{hhc+09,hhc+08}. 

The \textsc{tempo2} package provides a few ``plugin'' packages and routines suitable for GW analysis:
\begin{itemize}
\item{\textsc{GWsim.h}: a library that implements the mathematics described by Hobbs et al. (2009)\nocite{hjl+09}.  The routines in this library can be used to create new \textsc{tempo2} plugins easily.}
\item{\textsc{fake}: plugin that simulates pulsar TOAs given a timing model.}
\item{\textsc{GWsingle}: plugin that simulates a single non-evolving GW source and produces a file of site-arrival-times that can be used for subsequent processing.}
\item{\textsc{GWevolve}: plugin that simulates a single evolving GW source.}
\item{\textsc{GWbkgrd}: plugin that simulates a background of GW sources.}
\end{itemize}
A new plugin is currently under development that will either provide limits on the amplitude of the GW background or give the significance of any detection by searching for correlations in the timing residuals of multiple pulsars.

\section{The future}

It is likely that the signature of a GW background will be discovered in IPTA data sets shortly.  However, any such initial ``detection'' will only have a low significance (e.g. $\sigma \sim 2$ to $3$).  How this significance will increase with time will depend upon the intrinsic pulsar timing noise, the amplitude and spectral exponent of the background, whether new pulsars are discovered and whether observing systems are improved.  A very high significance detection, enabling detailed studies of the astrophysical objects that form the background and the physics of the GWs themselves (e.g. \cite{ljp08}), will require large numbers of pulsars to be observed with high precision. Various new telescopes are being developed that should detect a large number of new pulsars or be able to observe pulsars with high precision (e.g. the Australian Square Kilometre Array Pathfinder \cite{jbb+07,jtb+08}, MeerKAT \cite{bbjf09} and FAST \cite{nwz+06}). These telescopes are pathfinders to the Square Kilometre Array which should be fully operational by 2024.  This telescope should revolutionise pulsar and GW astronomy by allowing almost all pulsars in our Galaxy to be detected and GW detection using pulsar observations should become commonplace. 

\section{Conclusion}

Pulsar timing experiments can detect gravitational wave signals.  The most likely detectable source is an isotropic, stochastic background from a large number of coalescing supermassive binary black hole systems.  It is possible that a detection of such a background could be made with existing data, but it will require a further 5-10\,yr to have a definitive, highly significant detection.  It is expected that GW astronomy using pulsars will become a standard astronomical tool during the Square Kilometre Array era when the background will be analysed in detail and individual binary black-hole systems and burst GW sources will be detectable.

%It is possible that a `detection' of a stochastic background of gravitational waves will be made by pulsar timing experiments within the next few years.  However, any such `detection' will only have a low significance (i.e. $\sigma \sim 3$) and only lead to a poor estimate of the parameters needed to determine the amplitude and formation of the background.  Within 5-10 years it is likely that the background detection will be confirmed, but it will require observations using the next generation of radio telescopes (such as the Square Kilometre Array) to make detailed studies of ultra-low-frequency gravitational waves. 

%
\begin{acknowledgement}
This work is undertaken as part of the Parkes Pulsar Timing Array project, which is a collaboration between CSIRO Astronomy and Space Science, Swinburne University, the University of Texas, Brownsville and the University of California, San Diego.  The Parkes radio telescope is part of the Australia Telescope, which is funded by the Commonwealth of Australia for operation as a National Facility managed by CSIRO. GH thanks D. Yardley, J. Verbiest, R. Manchester and A. Sesana for comments on early drafts of this text.
\end{acknowledgement}

\bibliography{journals,modrefs,psrrefs,crossrefs}
\bibliographystyle{spmpsci}

\end{document}